\documentclass[a4paper]{jpconf}
\usepackage{graphicx}
\begin{document}
\title{Examination on SK atmospheric neutrino experiment by the computer experiment}
\author{A~Misaki$^1$, E~Konishi$^2$, N~Takahashi$^2$, Y~Minorikawa$^3$, V~I~Galkin$^4$, M~Ishiwata$^5$,  I~Nakamura$^5$ and M~Kato$^6$}
\address{$^1$ Advanced Research Institute for Science and Engineering, Waseda University, Tokyo, 169-0092 Japan}
\address{$^2$ Department of Electronic and Information System Engineering, Hirosaki University, Hirosaki, 036-8561 Japan}
\address{$^3$ Department of Science, Kinki-University, Higashi-Osaka, 577-8502 Japan}
\address{$^4$ Department of Physics, Moscow State University, Moscow, 119992 Russia}
\address{$^5$ Department of Physics, Saitama University, Saitama, 338-8570 Japan}
\address{$^6$ Kyowa Interface Science Co.Ltd., Saitama, 351-0033 Japan}
\ead{misakiakeo@air.ocn.ne.jp}

\begin{abstract}
We examine neutrino events occurring inside the SuperKamiokande (SK) detector
and those occurring outside the same detector using computer
simulations. We analyze the zenith angle distribution of Fully
Contained Events and show the method for the determination of the
incident neutrino by the SK group is unreliable. The analysis of the neutrino
events occurring outside the detector shows these events agree with
the Monte Carlo simulation without oscillation.
\end{abstract}

In our oral presentation at TAUP2005, we examined various topics in
atmospheric neutrinos in the SK and K2K experiments. However, in this
short paper, we are obliged to present our conclusions without
detailed derivations and verifications due to the 
restrictions on paper length.  We must
omit consideration of the K2K experiment here for the same reason.  Readers
are requested to refer to our (forthcoming) papers [1] for detailed
examinations.  

SK deals with the neutrino events occurring inside the
detector (Fully Contained Events and Partially Contained Events) and
events occurring outside the detector (Upward Through-Going Muon Events
and Stopping Muon Events).  In the analysis of the Fully Contained
Events and Partially Contained Events, the SK group adopts the assumption that
the direction of the incident neutrino is the same as that of the
emitted lepton. In order to avoid any misunderstanding, we reproduce this
assumption from the original SK papers:

[a] ``However, the direction of the neutrino must be estimated from the
reconstructed direction of the product of the neutrino interaction. In
water Cherenkov detectors, the direction of an observed charged lepton
is assumed to be the direction of the neutrino." [2]

[b] ``The direction of neutrino for FC single-ring sample is simply
assumed to be the same as the reconstructed direction of muon. Zenith
angle of neutrino is reconstructed as follows: 
$ cos \theta^{rec}_{\nu } = cos \theta _{\mu } $ (8.17) where 
$cos \theta _{\mu}$ and $cos \theta ^{rec}_{\nu }$ are cosine of 
the reconstructed zenith angle of muon and neutrino, respectively." [3]
 
In the analysis of neutrino events occurring inside the detector,
quasi elastic scattering (QEL) is the most important process, because QEL produces
a single ring event in Fully Contained Events which results in the most
clear cut events for analysis.  Here, we examine the influence of
neglecting the scattering angle of the charged lepton due to QEL
over the final zenith angle of the charged lepton.  Let $\theta_{s}$ 
and $\varphi _{s}$ be the scattering angle and the azimuthal angle,
respectively.  In the SK analysis, the zenith angle of the particles
concerned are utilized. Here, let $(\ell, m, n(=cos\theta_{\nu}))$, 
and $(\ell_{\mu }, m_{\mu }, n_{\mu }(=cos\theta_{\mu}) )$ be the
direct cosine of the incident neutrino and that of the emitted muon,
respectively. In this case, the relation between $(\ell, m, n)$
and $(\ell_{\mu }, m_{\mu }, n_{\mu })$ is given as follows:

\begin{equation}
\left(
\begin{array}{c}
\ell_{\mu } \\
m_{\mu }\\
n_{\mu }
\end{array}
\right)
=
\left(
\begin{array}{ccc}
 cos\theta cos\varphi & -sin\varphi & \ell \\
 cos\theta sin\varphi &  cos\varphi & m \\
 -sin\theta           &      0      & n
\end{array}
\right)
\left(
\begin{array}{c}
 sin\theta_{s} cos\varphi_{s} \\
 sin\theta_{s} sin\varphi_{s} \\
 cos\theta_{s}
\end{array}
\right)
\end{equation}

\noindent
where $cos \theta _{\nu }$ and $cos \theta _{\mu }$ denote the zenith
angle of the incident neutrino and that of the emitted muon,
respectively. Here, $\ell = sin\theta cos\varphi$ and $m = sin\theta
sin\varphi$.

It should be noticed that the influence of the azimuthal angle of the
events over their zenith angle can be very significant, particularly for 
inclined neutrinos and furthermore, the azimuthal angle may play an
important role in the ``transmutation" between Fully Contained Events
and Partially Contained Events in real analysis.

We obtained the zenith angle of the emitted muon for each sampled
incident neutrino as a function of zenith angle and energy. In Figure~1,
we give the scatter plot between 
$cos \theta _{\nu }$ and $cos \theta _{\mu }$.  The SK group utilize the relation (8.17) 
given above. It is, however, clear from Figure~1 that such an assumption
holds above 5\,GeV at most and that it does not hold at lower energies,
where most neutrino events of interest exist.

\vspace{-2mm}
\begin{figure}[h]
\begin{minipage}[t]{12pc}
\includegraphics[width=12pc]{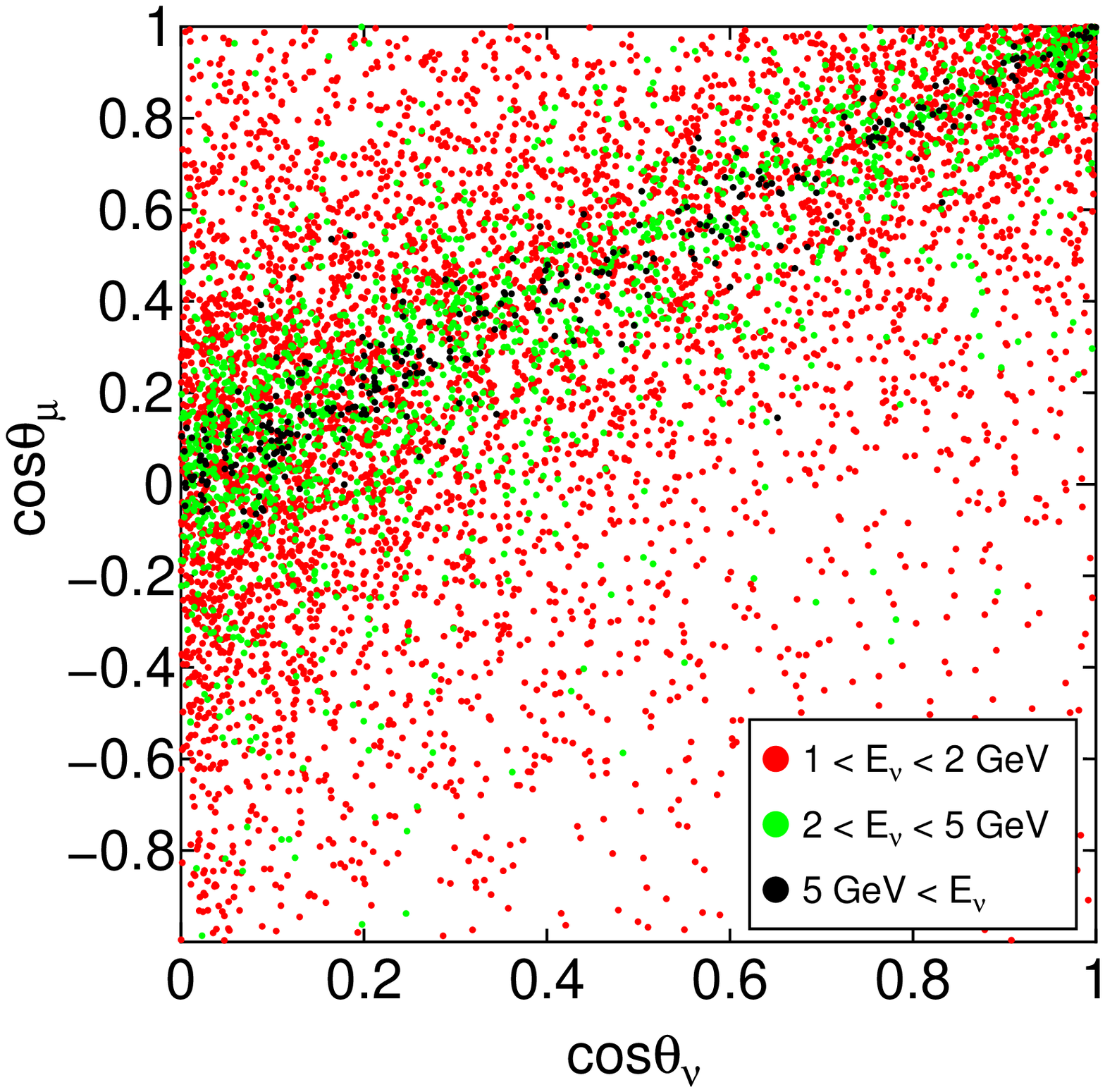}
\caption{\label{label}Correlation between $cos \theta _{\nu }$ and  $cos \theta _{\mu }$ for different neutrino energy ranges.}
\end{minipage}\hspace{1pc}%
\begin{minipage}[t]{12pc}
\includegraphics[width=12pc]{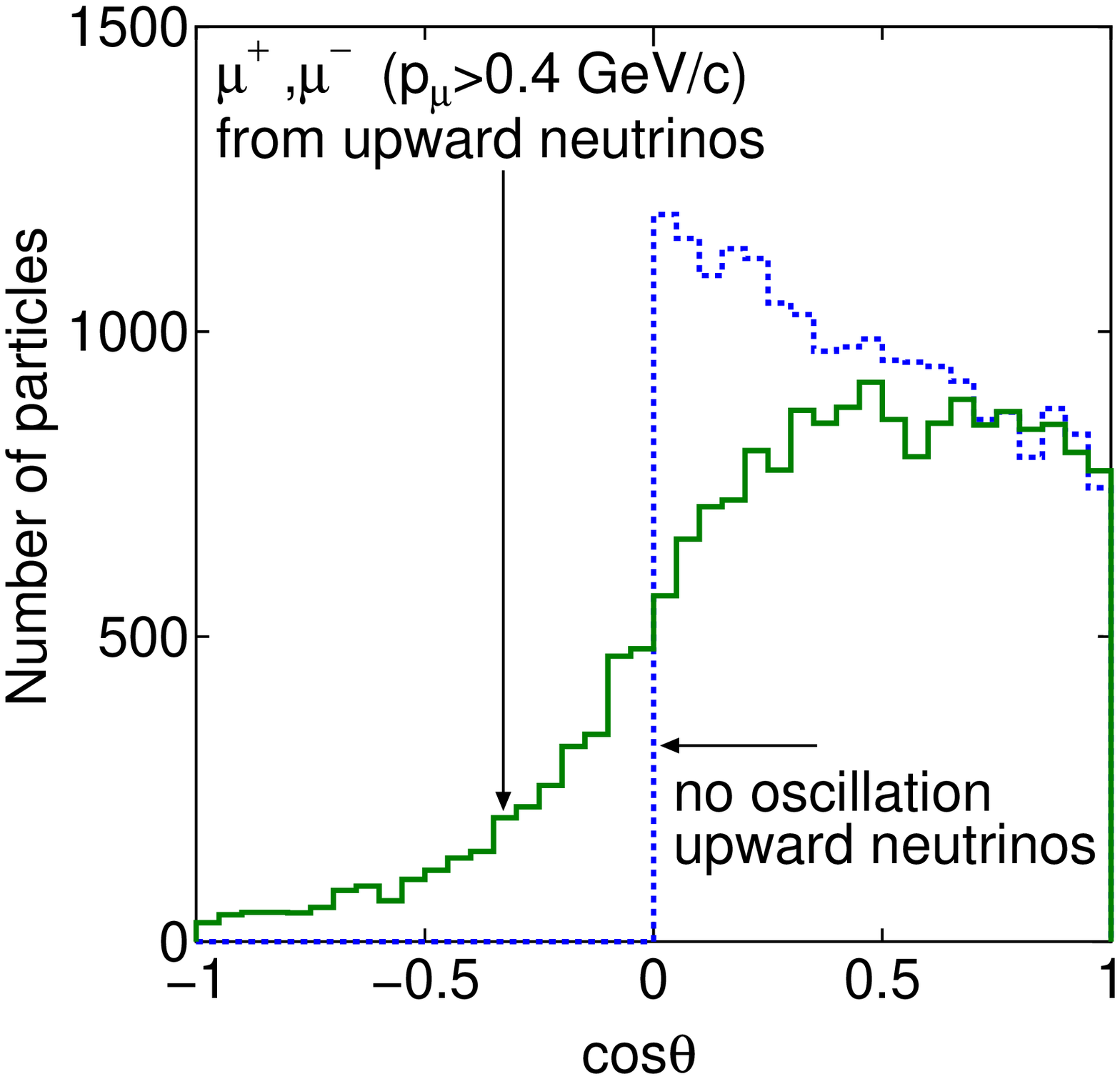}
\caption{\label{label}The relation between the zenith angle distribution of the incident neutrinos and corresponding one of the emitted muons.}
\end{minipage}\hspace{1pc}%
\begin{minipage}[t]{12pc}
\includegraphics[width=12pc]{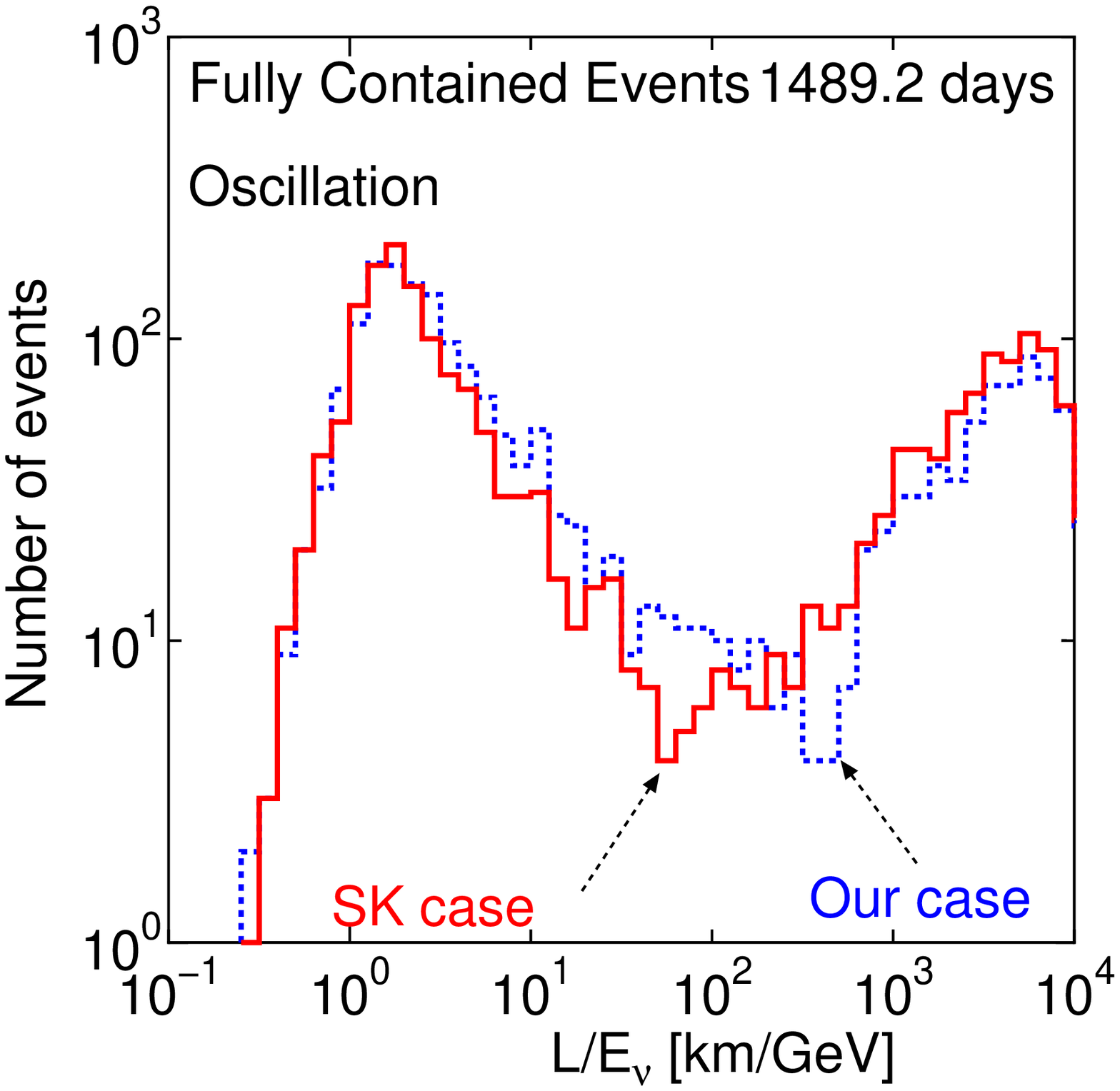}
\caption{\label{label}L/E distribution of muon-like single ring events due to QEL in the Fully Contained Events.}
\end{minipage}
\end{figure}

\vspace{-5mm} 
For the zenith angle distribution of upward neutrinos
whose energy spectrum in the SK detector is derived from the incident
neutrino energy spectrum at the surface of the Earth, we could
calculate the corresponding distribution of the emitted muons following Eq.(1).
In this case, we simulate the azimuthal angle as well as the
scattering angle of the emitted leptons exactly following the
probability function due to QEL. In Figure~2, the zenith angle
distribution of the emitted muons is given together with the
corresponding parent incident neutrinos.  According to the SK assumption,
the zenith angle distribution of the emitted muons is the same as the
distribution of the incident neutrinos. As illustrated, however, the
muon distribution is quite different from that of the incident neutrinos.
This is the reason why the determination of the incident neutrino direction
by the SK group is not reliable. Namely, we can not neglect the effect of the
scattering angle due to QEL in the analysis of events occurring inside
the detector. Also, we can not neglect the effect of the azimuthal
angle in the analysis. The SK group do not treat the effect of the
azimuthal angle in their analysis.

The SK group assert that they have found the signature of
neutrino oscillations in their analysis
of Fully Contained Events and Partially Contained Events [4]. 
It is, however, necessary to analyze Fully
Contained Events separately from Partially Contained
Events.

In Figure~3, we give the L/E distribution due to QEL in the presence
of oscillations with the SK parameters. In our case we exactly simulate
neutrino events, taking into consideration the effects of both back-scattering
and azimuthal angle, while in the SK case we adopt the SK group's
assumption that the incident neutrino and emitted muon have the same direction. 
In both our case and the SK case, we analyze Fully Contained
Events exclusively to eliminate ambiguities as much as possible,
whereas the real analysis by the SK group is done using a mixture of Fully Contained
Events and Partially Contained Events. In our case, we can clearly see
a dip around (301--501)~km/GeV, which roughly coincides with
the theoretically expected 530~km/GeV. In the SK case, the corresponding
dip occurs at (50.1--63.1)~km/GeV, which is smaller than 530~km/GeV by
one order of magnitude. This also shows that the SK group's assumption is not reliable.

In contrast to the analysis of neutrino events occurring
inside the detector, we can assume the direction of the emitted
lepton is the same as that of the incident neutrino in the analysis of
neutrino events outside the detector, namely, Upward Through-Going
Particles Events and 

\begin{minipage}[h]{12pc}
\begin{center}
\includegraphics[width=12pc]{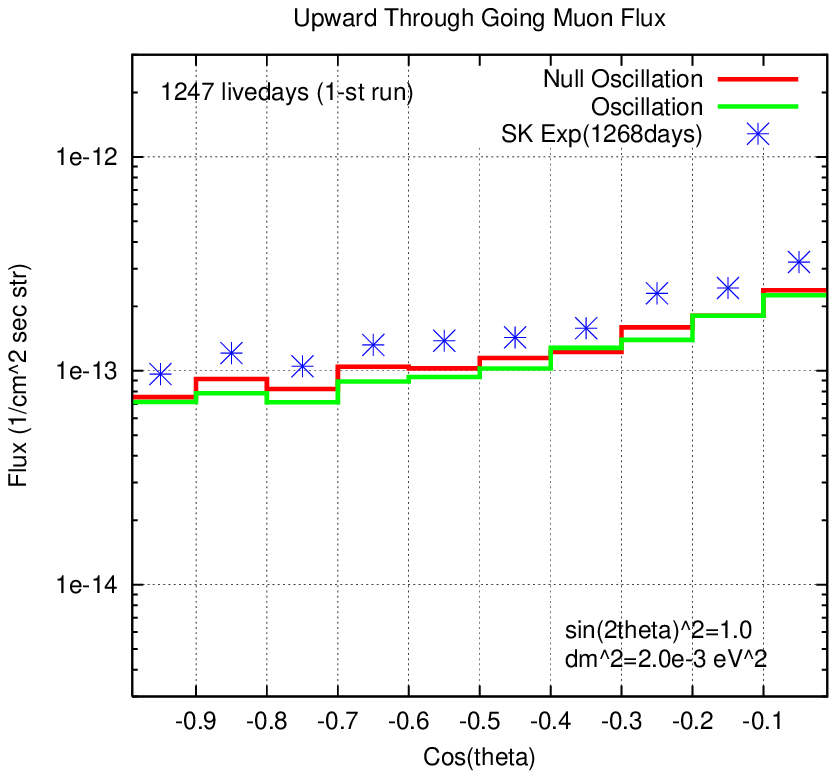}
\end{center}
\vspace{-1mm}
\textbf{Figure 4.} Zenith angle distribution of Upward Through Going Muon Events. 
\end{minipage}\hspace{1pc}%
\begin{minipage}[h]{12pc}
\begin{center}
\includegraphics[width=12pc]{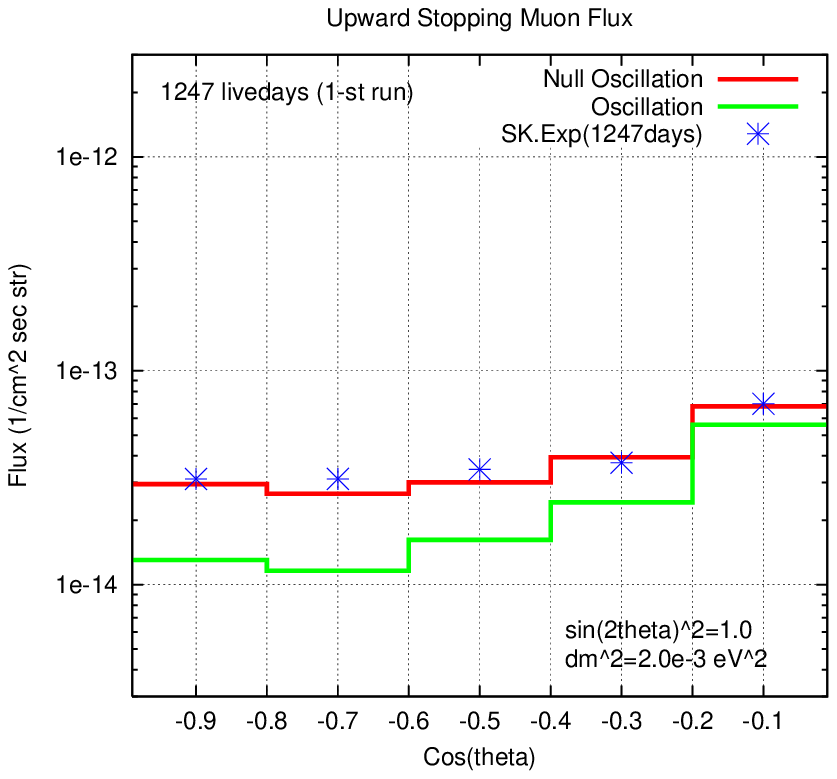}
\end{center}
\vspace{-1mm}
\textbf{Figure 5.} Zenith angle distribution of Upward Stopping Muon Events.
\end{minipage}\hspace{1pc}%
\begin{minipage}[h]{10.8pc}
\vspace{1mm}
Stopping Particles Events may be generated exclusively by high energy muons due to deep inelastic scattering, 
where their scattering angles could be neglected. \\
\ \ We analyze physical events, taking into account the effect of the range fluctuation exactly, while SK use the 
average range of the muon concerned, being helped by an ``observation probability". This does 
\vspace{2mm}
\end{minipage}
not reflect the real situation of limited statistics. In Figures~4 and 5, 
we give the zenith angle distributions of Upward
Through-Going Muon Events and Stopping Muon Events, taking into
account all fluctuation effects exactly, and compare our results with
the experimental data by SK directly. The SK experimental data are
consistent with our results without oscillation [5].

Thus, we find 
the analysis of atmospheric neutrinos by the SK group
does not provide  
definite evidence for the existence of the neutrino oscillations.

\end{document}